\documentclass[a4paper, nocompress]{spie}  
\usepackage[]{graphicx}
\usepackage{textcomp}

\title{The ASTRA project: a doorway to future astrometry} 

\author{M. Gai\supit{a}, Z. Qi\supit{b}, M.G. Lattanzi\supit{a}, 
B. Bucciarelli\supit{a}, D. Busonero\supit{a}, M. Crosta\supit{a}, F. 
Landini\supit{a},  S. Liao\supit{b}, H. Luo\supit{b}, G. Mana\supit{c}, 
R.A. M\'endez\supit{d}, M. Pisani\supit{c}, A. Riva\supit{a}, C. San Martin 
Luque\supit{d}, C.P. Sasso\supit{c}, Z. Tang\supit{b}, A. Vecchiato\supit{a}, 
Y. Yu\supit{b}
\skiplinehalf
\supit{a}INAF - Osserv. Astrofisico di Torino, V. Osservatorio, 20, I-10025 Pino Torinese (TO), Italy \\
\supit{b}Shanghai Astronomical Observatory, CAS, 80 Nandan Rd, Shanghai 200030, China\\
\supit{c}INRiM - Ist. Naz. di Ricerca Metrologica, Str. delle Cacce, 91, I-10135 Torino (TO), Italy \\
\supit{d}Departamento de Astronomia, Universidad de Chile, Casilla 36-D, Santiago, Chile\\
}
\authorinfo{Further author information: E-mail: mario.gai@inaf.it, Telephone: +39 011 8101 943, www.oato.inaf.it; E-mail: zxqi@shao.ac.cn, Telephone: +86 21 3477 5203, english.shao.cas.cn }

\begin{document} 
  \maketitle 
\begin{abstract}
Astrometric Science and Technology Roadmap for Astrophysics (ASTRA) is a bilateral cooperation 
between China and Italy with the goal of consolidating astrometric measurement concepts and 
technologies. 
In particular, the objectives include critical analysis of the Gaia methodology and performance, 
as well as principle demonstration experiments aimed at future innovative astrometric 
applications requiring high precision over large angular separations (one to 180 degrees). 
Such measurement technologies will be the building blocks for future instrumentation focused 
on the ``great questions" of modern cosmology, like General Relativity validity (including 
Dark Matter and Dark Energy behavior), formation and evolution of structure like proto-galaxies, 
and planetary systems formation in bio compatibles environments.
We describe three principle demonstration tests designed to address some of the potential 
showstoppers for high astrometric precision experiments. 
The three tests are focused on the key concepts of multiple fields telescopes, astrometric 
metrology and very fine sub-pixel precision (goal: $<1/2000$ pixel) in white light.
\end{abstract}



\section{Introduction}
\label{sec:introduction}  

This paper describes the scope and the initial activity of the ASTRA 
(Astrometric Science and Technology Roadmap for Astrophysics) project. 
ASTRA is a collaboration between the Italian Institute for Astrophysics 
(INAF), in particular the astrometric team at the Astrophysical Observatory 
of Turin (OATo), and the Shanghai Astronomical Observatory (SHAO, Chinese 
Academy of Sciences - CAS), under the auspices of their respective Ministries, 
in the framework of the Scientific and Technological Cooperation 
Agreement between China and Italy.  
The project, started in late 2019, builds on a long collaboration between 
the teams in the field of high precision astrometry, in particular on 
materialization of the absolute reference frame (APOP, \cite{Qi2015}), 
the astrometry mission proposal GAME (Gravitation Astrometric Measurement 
Experiment) to ESA \cite{GameExpAstr2012, Gai2012GAME, Gai_GAME_2014},
its 2015 AGP (Astrometric Gravitation Probe) rendition \cite{Gai15Met}, 
and the ongoing experience on Gaia \cite{GaiaDR1_2016, GaiaDR2Summary, 
GaiaDR2Astrometry}. 
ASTRA also aims at fostering and consolidating national and international 
collaborations on astronomical and interdisciplinary research related to 
astrometric advancement. 

\noindent 
The ``great questions" of modern Cosmology, summarising e.g. ESA's Cosmic Vision 
``Grand Themes", are: 
\begin{enumerate}
\item Validity of General Relativity (GR) at all scales, hence the nature of Dark 
Matter and Dark Energy and their relationship with baryonic structures;
\item structure formation and evolution, from primordial galaxies to the Milky Way; 
study of distant and local gravitational wave sources; 
\item formation of planetary systems (including the Solar System) and life 
supporting environments. 
\end{enumerate}

The main goal of ASTRA is the deployment of shared experience on enabling astrometric 
technologies aimed at future large angle, high precision measurements. 
The activity is based on synergy of science and technology: we will build on our GR 
models for the Gaia mission, to consolidate science requirements for astrometric 
technology  \cite{Lattanzi2012, Crosta19}: measurement at $<1\ \mu as$ implies simulation 
accuracy in the range $0.1$ to $0.1 \,\mu as$ to ensure control of model errors. 
The long-term objective is the development of innovative astronomic instrumentation 
reliably operating in the micro-arcsecond ($\mu as$) precision range or better, also 
including relativistic experiments, as the above mentioned GAME/AGP, or a novel 
astrometric instrument concept presented in another contribution to this meeting 
(RAFTER, \cite{RAFTER_SPIE_20}). 
This will be achieved through modelization, simulation and laboratory experiments, 
addressing some basic implementation aspects of astrometric measurements, described 
below. 

The ASTRA activity follows an historical path started with Hipparcos 
\cite{Hipparcos97}, continues with Gaia, and aims toward future 
groundbreaking experiments on astrophysics and fundamental physics, e.g. 
JWST \cite{Fortenbach20}, NEAT \cite{Malbet12}, Theia \cite{Malbet16}, Gaia 
extensions to higher precision or other spectral bands, in particular NIR 
\cite{GaiaNIR18}, and others \cite{Hahn18, RAFTER_SPIE_20}. 
The main output of ASTRA will be a white paper summarising a 
roadmap toward future high precision astrometry.

Modern astrophysical instrumentation aims at photon limit precision. 
Current missions (like Gaia and Euclid) probably exploit present 
technological limits on passive stabilization, while future ones 
(like LISA), based on distributed instruments, intrinsically rely  
on advanced metrology concepts. 
Observing fairly bright stars ($V = 8-10\ mag$) in the visible range, 
a $1\ m$ class telescope can achieve a nominal $\mu as$ level precision 
in a fairly short exposure time, of order of one hour. 
In practice, comparable accuracy, i.e. systematic error control, is 
hard to achieve. 
This is related to knowledge of the instrument response throughout the 
measurement, hence calibration.
Moderate variation of the instrument response over the field of view, 
observing bandwidth, and in time, are of course beneficial toward good 
modeling and reliable estimation of instrumental parameters, therefore 
usually stringent requirements are issued on imaging quality and 
stability. 

\noindent 
Among the main challenges considered for (sub-)$\mu as$ astrometry, 
relevant sources of systematic errors are: 
\begin{enumerate}
\item the field variation of telescope optical response; 
\item the variation of electro-optical response over detectors; 
\item ``Cosmic noise", i.e. the (real or apparent) variability of 
individual astronomical objects. 
\end{enumerate}
Moreover, the capability of large angle measurement (a few to several ten degrees) 
is a crucial requirement either in the context of global astrometry (with the 
approach of Hipparcos and Gaia), or looking for adequate, typically bright, 
reference sources. 
\\ 
The ASTRA study includes a critical review of reference science goals, deriving 
a detailed definition of scientific requirements, also taking advantage of the 
lessons learned from Gaia, and setting the specifications for relevant aspects 
of future applications. 

As a possible mitigation strategy with respect to the above challenges, 
we investigate the concept of an annular field telescope, RAFTER: Ring Astrometric 
Field Telescope for Exo-planets and Relativity \cite{RAFTER_SPIE_20}. 
It features highly uniform optical response over a large focal plane area, 
thus providing favourable characteristics with resect to the possibility of 
averaging out detector and source variability over a set of observations. 
A specific application for Fundamental Physics, with the addition of a custom 
coronagraphic system, is represented by the current Astrometric 
Gravitation Probe design\cite{Gai2020AGP}. 

\subsection{Technological objectives}
\label{ssec:objectives}  
The above critical aspects have been assessed\cite{Shao19, Malbet20} 
in the literature; 
in the context of ASTRA, three specific concepts will be tested as potential 
contributions toward systematics control. 
They are described in the following sections, and can be summarized as: 
\begin{enumerate}
\item multiple line-of-sight (LOS) telescopes: 
study, optimization and prototyping of a three LOS telescope, compatible with metrology, 
to materialize the bidimensional angular gauge onto the sky;
\item astrometric metrology: a high precision metrology device strictly embedded 
in the telescope design, for the real-time estimation\cite{2013PASP..125.1383G} 
of systematic error induced by external sources, using BAM-like\cite{Riva2014} 
fringes; 
\item image centering experiment: design, simulation, implementation and data 
analysis of a setup for high precision location of images on CCD and CMOS 
detectors, down to  $1<2000$ pixel, through calibration and differential techniques. 
\end{enumerate}

\section{Experiment Description}
\label{sec:experiment}  
%
%
The first period of activity is mostly focused on implementation of the Image 
Centering Experiment described in Sec.\,\ref{sec:ICE}, whereas the other aspects 
are investigated at the moment mostly at the level of design and simulation. 

\subsection{Image Centering Experiment}
\label{sec:ICE}
The principle is similar to that implemented in Gai et al. (2001)\cite{GaiAA2001}, 
and consists in the simplest representation of an imaging instrument, i.e. a camera 
fed by a doublet, observing a simulated stellar field. 
The setup also has some educational value, e.g. allowing students to get hands-on 
experience on the problem without consuming precious telescope time. 
\\ 
The expected result on location uncertainty are described in the literature 
\cite{GaiAA2001,Mendez2013,Gai2017PASP}, and its relationship with the 
theoretical Cram{\'e}r-Rao limit is also discussed in other  papers\cite{Echeverria16,Espinosa18}. 
In practice, the image photocenter, in low background conditions, is affected 
by errors depending on geometric factors related to the PSF shape, and scaling 
inversely to the SNR, at least for bright sources. 
For a Gaussian PSF\cite{Mendez2013} centered in $x_c$, with Gaussian width $\sigma$: 
\begin{equation}
\Phi(x;x_c, \sigma) = 
\frac{1}{\sqrt{2\pi}\sigma} 
\exp\left[-\frac{\left(x-x_{c}\right)^{2}}{2\sigma^{2}}\right] \, . 
\end{equation}
the location uncertainty $\sigma_c$ is 
\begin{equation}
\sigma_c \ge \frac{\sigma}{SNR} \, . 
\label{Eq:PrecSNR}
\end{equation}
Similar expressions can be achieved for large classes of PSF shapes, replacing 
the Gaussian width $\sigma$ with a 
geometric factor, expressed e.g. as a function of the diffraction 
parameter $\lambda/D$, to be computed\cite{Lindegren1978,GaiPASP1998} 
according to imaging performance and measurement conditions. 
Although Eq.\ \ref{Eq:PrecSNR} is rather general, the dependence on source 
magnitude, readout noise, background and actual PSF shape is totally 
implicit; more detail can be found in the above literature and references 
therein. 

In the framework of our collaboration, two versions of the same concept have been 
realized, in each Institute's labs. They are similar but not identical, in order 
to cover a wider range of possible choices. Measurements will be taken with both 
CMOS and CCD sensors, in order to evidence some of the characteristics of each 
technology. 
In both versions, since the source images are significantly smaller than the detector 
area, it is possible to perform response tests over large fractions of the detector 
sensitive area simply by offsetting the camera, i.e. implementing a sort of dithering 
strategy in the lab. 
The simulated stellar field takes therefore the role of conventional test patterns 
in commercial imaging applications. 
The SHAO version also features comparable brightness of all artificial stars, 
whereas the OATo version generates sources on a range of about three 
magnitudes.

\begin{figure}[ht]
  \begin{center}
    \includegraphics[width=0.75\textwidth]{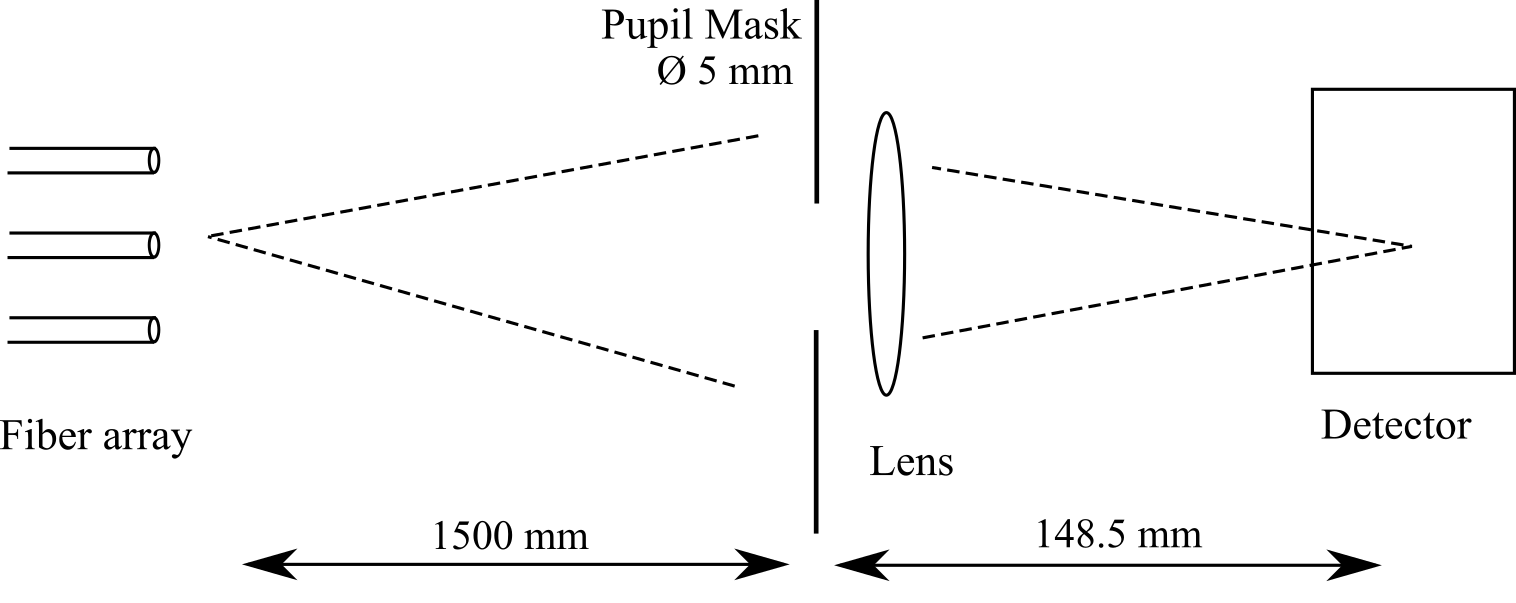}
    \caption{\label{fig:OpticalLayoutSH} Optical layout of SHAO image centering experiment 
    (not to scale).}
  \end{center}
\end{figure}

\subsubsection{SHAO implementation}
\begin{figure}[ht]
  \begin{center}
    \includegraphics[width=0.75\textwidth]{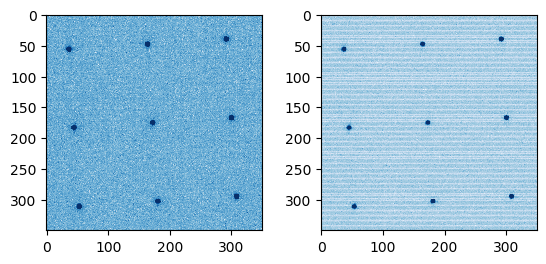}
    \caption{\label{fig:FrameSH} Field image of SH location experiment, log scale; 
    normal frame (left), and frame affected by excess spatial noise (right). }
  \end{center}
\end{figure}

The SHAO version adopts an array of fibers arranged in a regular grid, reimaged 
onto a Sony QHY4040 CMOS camera (front illuminated sensor, $4k \times 4k$ 
format, pixel size 9\,$\mu m$), 
according to the scheme depicted in Fig.\ \ref{fig:OpticalLayoutSH}. 
The resulting images occupy a limited detector region (about $350 \times 350$\ pixels), 
and two sample frames are shown in Fig.\ \ref{fig:FrameSH} (log scale). 
The spots are oversampled, in order to reduce the sensitivity to pixel response 
variation; this also provides a larger dynamic range. 
The image quality variation over the field is negligible. 

\begin{figure}[ht]
  \begin{center}
    \includegraphics[width=0.75\textwidth]{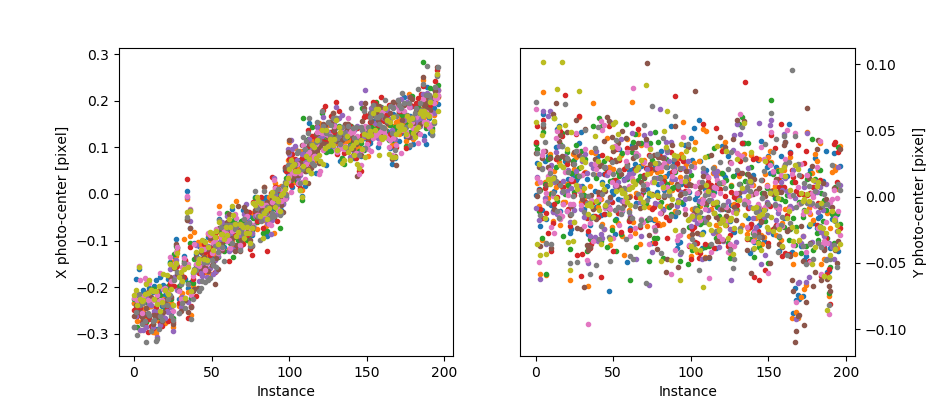}
    \caption{\label{fig:SH_XYcentsRaw} Raw photo-center dispersion, along X (left) 
    and Y (right) [SH].}
  \end{center}
\end{figure}
\begin{figure}
  \begin{center}
    \includegraphics[width=0.75\textwidth]{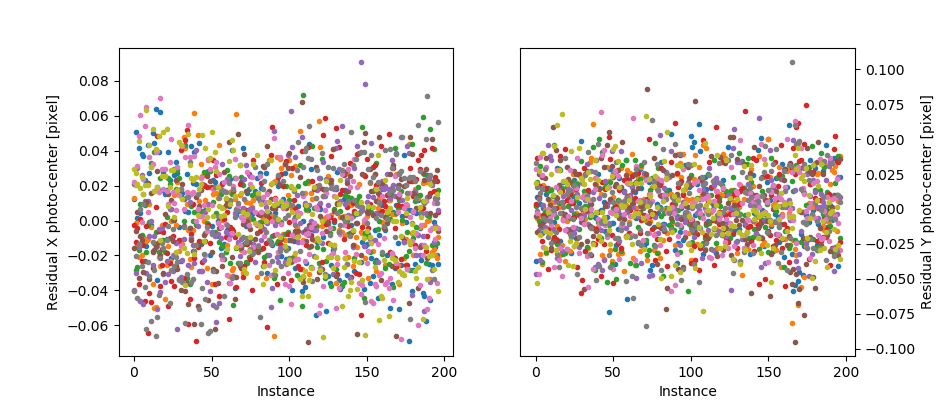}
    \caption{\label{fig:SH_XYcentsRed} Residual photo-center dispersion, along X (left) 
    and Y (right), after common mode removal [SH].}
  \end{center}
\end{figure}
\begin{figure}
  \begin{center}
    \includegraphics[width=0.75\textwidth]{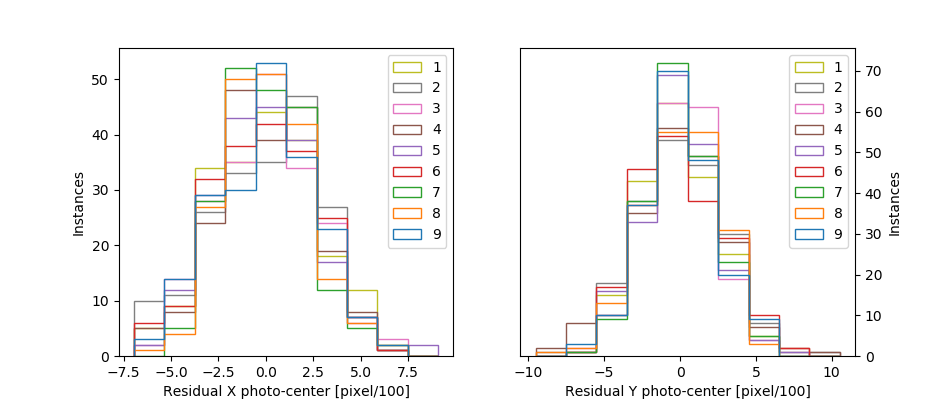}
    \caption{\label{fig:SH_XYRedHist} Residual photo-center histogram, along X (left) 
    and Y (right), after common mode removal [SH].}
  \end{center}
\end{figure}

The pupil size is set to approximately 2 mm, and the light 
source, very bright, required neutral filters required to 
prevent saturation even with the shortest exposure 
time [1 ms] made available by the camera control electronics. 
Alignment and focusing are undergoing optimization.
\\ 
The image quality is dominated by the resolved fiber core, providing a Gaussian spot 
with size $\sigma \simeq 6\,$pixel. 
The fiber array is not aligned with the detector axes.
In the preliminary integration and alignment test, four short sequences of full 
frames were taken, at the maximum rate of 10 frames/s. 

The preliminary evaluation of the images evidenced a comparably large spatial 
noise ($\sim 90$\ ADU), not justified by the background level ($\sim 1700$\ ADU) 
and nominal readout noise ($\sim 4$\ electrons). 
The trouble is likely due to environmental EM noise, whose source is under 
investigation. 
In spite of such significant noise, the image location is reasonably good. 

The location algorithms used to process the image set are a simple center-of-gravity 
(COG), and gaussian fit. 
They provide quite similar results, as expected from the Gaussiam image shape 
and high SNR conditions, but the latter provides of course somewhat lower 
sensitivity to the spatial noise excess experienced on some frames. 
\\ 
Dispersion, shown in Fig.\ \ref{fig:SH_XYcentsRaw}, is mainly due to common mode 
drift of order of 0.1 pixel, mostly over the X axis. 
Removal of the common mode significantly abates the dispersion, to the order 
of 0.01 pixel, somewhat larger on the Y axis. 
The residual photo-center dispersion is shown as a scatter plot in Fig.\ \ref{fig:SH_XYcentsRed}, and as a set of stacked histogram in 
Fig.\ \ref{fig:SH_XYRedHist}. 
All simulated star, with approximately the same brightness and signal level, 
have comparable dispersion. 

The statistics on photo-center dispersion is reported in Table\ \ref{tab:SH_LocStat}, 
in which the standard deviation of initial source positions in frame coordinates is 
reported as X(Y)\ Raw. 
The RMS coordinate residual after common mode subtraction, corresponding to fluctuation 
with respect to the overall motion of the set of stars, is reported as X(Y)\,Res, and 
can be compared with the corresponding Cram{\'e}r-Rao limiting value X(Y)\,CRL in 
the subsequent line; the ratio between the two values is reported as X(Y)\,Rel. 

\begin{table}[ht]
\caption{Location performance statistics on X and Y photo-center. 
Raw: uncorrected estimates; Res.: common mode subtracted; CRL: Cram{\'e}r-Rao 
limit; Rel.: ratio between measured dispersion and Cram{\'e}r-Rao limit [SH]. } 
\label{tab:SH_LocStat}
\begin{center}       
\begin{tabular}{l|c|c|c|c|c|c|c|c|c} 
 & 1 & 2 & 3 & 4 & 5 & 6 & 7 & 8 & 9 \\ 
\hline
X Raw [px] &  0.1332  &  0.1495  &  0.1470  &  0.1579  &  0.1590  &  0.1541  &  0.1311  &  0.1634  &  0.1308 \\
Y Raw [px] &  0.0295  &  0.0316  &  0.0287  &  0.0309  &  0.0295  &  0.0375  &  0.0286  &  0.0303  &  0.0314 \\
\hline
X Res [px] &  0.0256  &  0.0225  &  0.0216  &  0.0268  &  0.0262  &  0.0253  &  0.0264  &  0.0285  &  0.0267 \\
X CRL [px] &  0.0191  &  0.0196  &  0.0207  &  0.0211  &  0.0217  &  0.0215  &  0.0197  &  0.0202  &  0.0226 \\
X Rel  &  1.3443  &  1.1450  &  1.0453  &  1.2724  &  1.2040  &  1.1754  &  1.3410  &  1.4109  &  1.1807 \\
\hline
Y Res [px] &  0.0245  &  0.0246  &  0.0226  &  0.0272  &  0.0232  &  0.0289  &  0.0229  &  0.0273  &  0.0251 \\
Y CRL [px] &  0.0189  &  0.0213  &  0.0210  &  0.0206  &  0.0227  &  0.0218  &  0.0196  &  0.0204  &  0.0225 \\
Y Rel  &  1.2973  &  1.1520  &  1.0756  &  1.3220  &  1.0207  &  1.3280  &  1.1671  &  1.3372  &  1.1171 \\
\end{tabular}
\end{center}
\end{table} 

The experimental dispersion X(Y)\,Res ranges between a few percent and a few ten 
percent above the theoretical best performance X(Y)\,CRL; the variation appears 
to be due to (a) actual image shape not exactly Gaussian, (b) approximate SNR 
estimation, and (c) unmodelled effects of spatial noise. 
\\ 
The results match the expectations for the observation SNR ($\sim 300$) and 
Gaussian image profile shape ($\sigma \simeq 6$\,pixels), within the simple 
assumptions applied. 
The location dispersion, of order of 1/50 pixel, correspond to about 1/500 of 
the actual image width, due to the oversampling.

\subsubsection{OATo implementation}
\begin{figure}[ht]
  \begin{center}
    \includegraphics[width=0.9\textwidth]{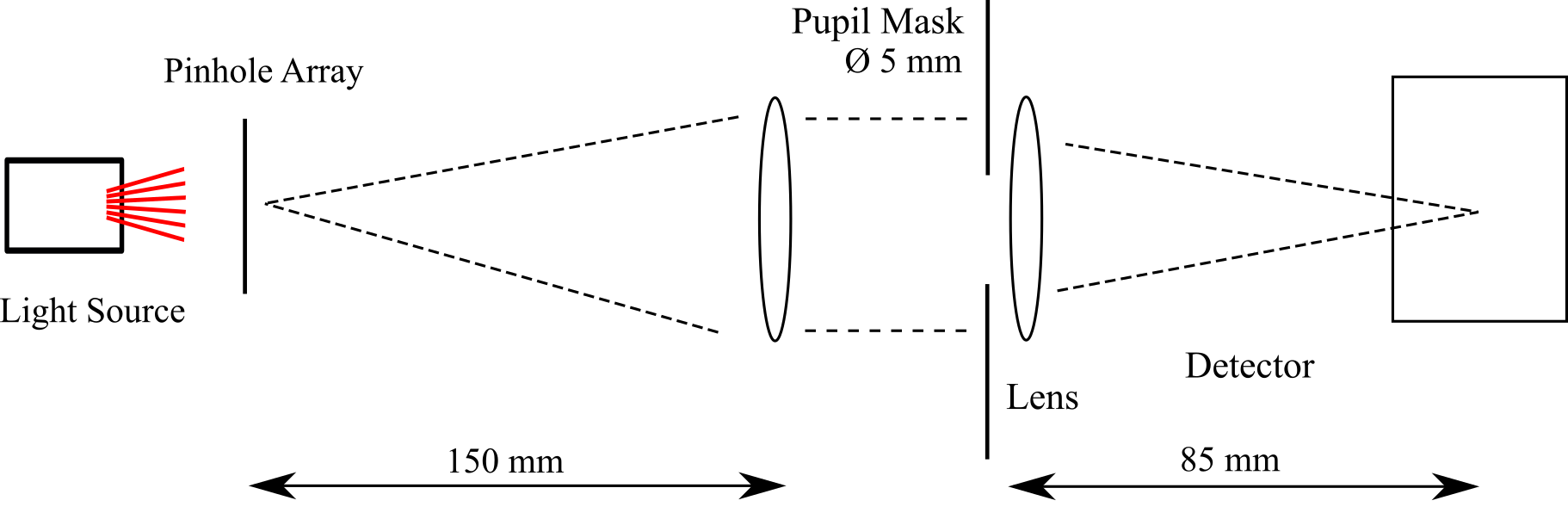}
    \caption{\label{fig:OpticalLayoutTO} Optical layout of OATo location experiment 
    (not to scale).}
  \end{center}
\end{figure}

\begin{figure}[ht]
  \begin{center}
    \includegraphics[width=0.42\textwidth]{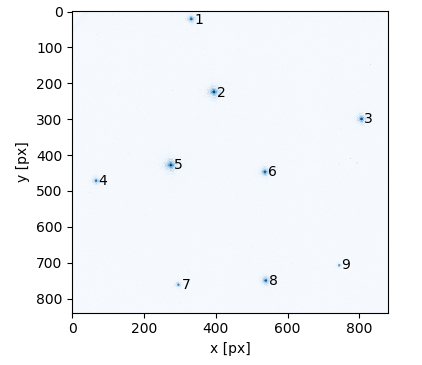}
    \includegraphics[width=0.42\textwidth]{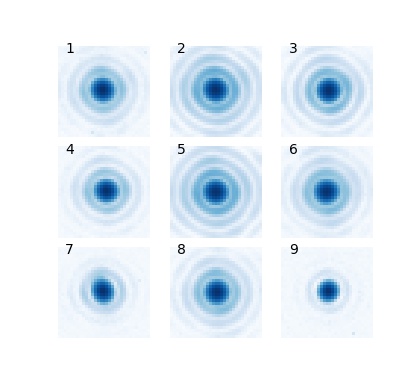}
    \caption{\label{fig:FrameTO} Field image (log scale) of (left), 
    and zoomed view of the nine spots (right).}
  \end{center}
\end{figure}

The OATo version uses a lab LED source (Thorlabs M625L4, nominal wavelength 625\,nm, 
bandwidth 17\,nm) and a custom array of pinholes irregularly spaced, reimaged 
onto a Kiralux CS235MU CMOS Camera ($1,920 \times 1,200$ format, pixel size 
5.86\,$\mu m$), 
according to the scheme depicted in Fig.\ \ref{fig:OpticalLayoutTO}. 
The pinhole array is the same used in activity for a previous 
publication\cite{Gai2001A&A}. 
The resulting images occupy a limited detector region (about $800 \times 850$\ pixels), 
and a sample frame is shown in Fig.\ \ref{fig:FrameTO} (left); the right panel shows a 
zoomed version of the nine spots. 
The images are nearly diffraction limited, since up to four diffraction rings are 
clearly visible, on the brightest sources, around the central peak. 
The image quality variation over the field is negligible. 

The pupil size is again $\sim 2\,mm$, and the source is sufficiently bright to 
ensure a signal level reaching a reasonable fraction of the full dynamic range 
with $10\,ms$ exposures; due to the different illumination of the pinholes, the 
photon flux of the simulated sources ranges over about one order of magnitude. 
\\ 
The background level, after partial optimization, is $\sim 2$\,ADU, with much lower 
spatial noise (a few ADU) than experienced in the previous setup, thus comparing well 
with the nominal RON ($\sim 1$\,ADU). 
The analysis below refers to a set of 100 frames collected in stable environment 
conditions. 

The location algorithms (apart COG for preliminary tests) are unweighted least 
square (LSQ) and maximum likelihood (ML) fit to the nominal polychromatic PSF 
of an ideal circular aperture, estimating PSF radius, photo-centre position 
and background. 
In the ML framework, we also estimate the expected error (MLEE), using expressions 
consistent with those derived e.g. in the 1D case in Gai et al., 
2017\cite{Gai2017PASP}.
\\ 
Since the camera dynamic range is limited to 12 bits, it was not possible to 
increase significantly the analogic integration; we therefore explore the option 
of digital integration by stacking the frames four at a time, mimicking the 
results of a fourfold increase in the exposure time. 
In the best case of purely random noise, with low systematics, this should improve 
the measurement precision b a factor two. 
The processing results from individual and stacked frames are labelled by ``I" and 
``S", respectively. 

\begin{table}[ht]
\caption{Location performance statistics on X and Y photo-center, using 
the Maximum Likelihood estimator on individual frames (I) and on 4-by-4 stacked 
images (S) [TO]. 
Raw: uncorrected estimates; Res.: common mode subtracted; MLEE: Maximum Likelihood 
Expected Error. Units: 1/100 pixel. } 
\label{tab:TO_LocStat}
\begin{center}       
\begin{tabular}{l|c|c|c|c|c|c|c|c|c} 
 & 1 & 2 & 3 & 4 & 5 & 6 & 7 & 8 & 9 \\ 
\hline
X I Raw &   2.093 &   1.908 &   2.195 &   1.736 &   1.783 &   2.022 &   2.014 &   1.969 &   2.387   \\
Y I Raw &   2.523 &   2.393 &   2.582 &   2.055 &   2.275 &   2.410 &   2.436 &   2.193 &   2.744   \\
\hline
X I Res  &   0.853 &   0.507 &   1.072 &   1.005 &   0.531 &   0.563 &   1.113 &   0.657 &   1.684   \\
X MLEE   &   0.644 &   0.284 &   0.327 &   0.415 &   0.278 &   0.494 &   1.144 &   0.503 &   1.673   \\
\hline
Y I Res  &   0.986 &   0.617 &   0.706 &   0.814 &   0.478 &   0.494 &   1.513 &   0.873 &   1.954   \\
Y MLEE   &   0.642 &   0.294 &   0.322 &   0.444 &   0.289 &   0.509 &   1.032 &   0.499 &   1.545   \\
\hline
X S Raw &   1.071 &   1.058 &   1.333 &   0.823 &   0.958 &   1.057 &   0.939 &   1.089 &   1.355   \\
Y S Raw  &   2.083 &   1.984 &   2.045 &   1.705 &   1.887 &   1.951 &   1.600 &   1.702 &   1.626   \\
\hline
X S Res &   0.336 &   0.266 &   0.734 &   0.479 &   0.374 &   0.183 &   0.591 &   0.290 &   0.846   \\
Y S Res &   0.489 &   0.274 &   0.344 &   0.396 &   0.274 &   0.294 &   0.617 &   0.466 &   0.994   \\
\end{tabular}
\end{center}
\end{table} 

\begin{figure}[ht]
  \begin{center}
    \includegraphics[width=0.4\textwidth]{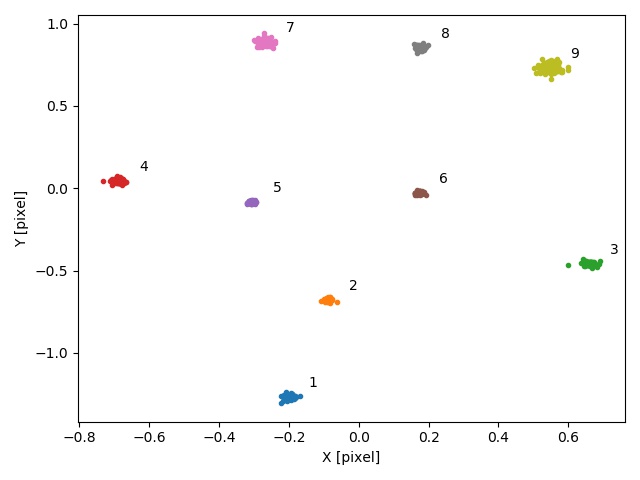}
    \includegraphics[width=0.4\textwidth]{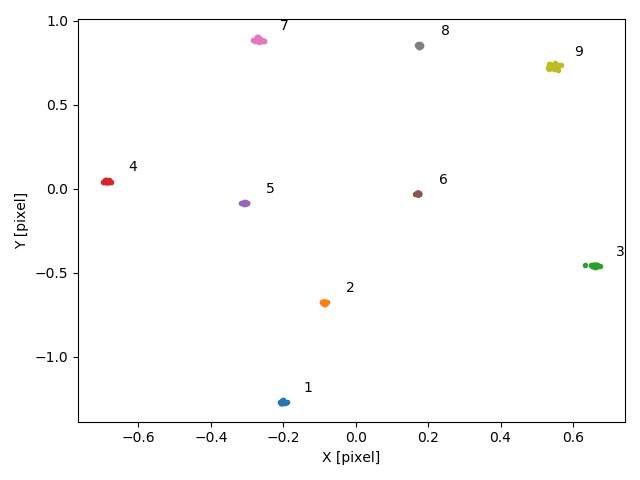}
    \caption{\label{fig:FrameTO_Disp} Scatter plot of star location dispersion around the nominal position (not to scale) for individual (left) and stacked (right) frames [TO].}
  \end{center}
\end{figure}

As in the case of the SHAO implementation, the standard deviation of raw photo-centre 
estimates, listed in Table\,\ref{tab:TO_LocStat} and labelled ``Raw", 
appears to be dominated by a significant common mode motion, somewhat reduced with minor 
improvements on mounting mechanics to the order of 0.02 pixel. 
After common mode subtraction, the dispersion of residuals (``Res") is reduced by 
a significant factor, below the 0.01 pixel level. 
\\ 
A representation of the different level of location error is shown in 
Fig.\,\ref{fig:FrameTO_Disp}, where residual photo-centre values are shown as 
displacement with respect to the nominal frame positions (not to scale), 
respectively for individual (left) and stacked (right) images. 
The apparent size of the cloud of points is significantly reduced in the latter case. 

\begin{figure}[ht]
  \begin{center}
    \includegraphics[width=0.7\textwidth]{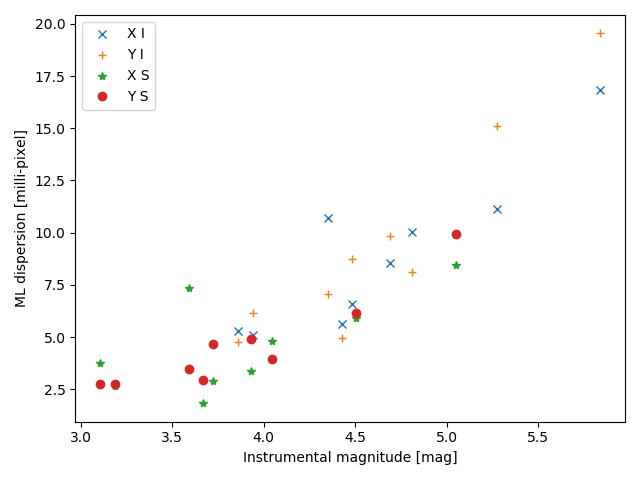}
    \caption{\label{fig:DispMagTO} Location dispersion of individual and stacked images 
    vs. instrumental magnitude [TO].}
  \end{center}
\end{figure}

Image stacking actually provides an improvement on photo-centre dispersion by a factor 
close to the expected value two, although the difference among sources is large. 
Also, the dispersion of residuals from individual frames compares well with the 
predicted MLEE, although the difference between theory and experiment 
is quite different for different sources, ranging from few percent to 
a factor two. 
In some cases, the experimental spread is even slightly lower than the MLEE. 
This appears to be related to the unmodelled sources of error, e.g. the crude 
image model, field dependent perturbations, and the limited statistics. 
\\ 
The location uncertainty is shown in Fig.\,\ref{fig:DispMagTO} as a function of the 
instrumental magnitude, including the results from both individual and stacked frames. 
The overall trend is roughly in agreement with the nominal behaviour suggested by 
theory, i.e. a linear dependence. 

\begin{figure}[ht]
	\begin{center}
		\includegraphics[width=0.7\textwidth]{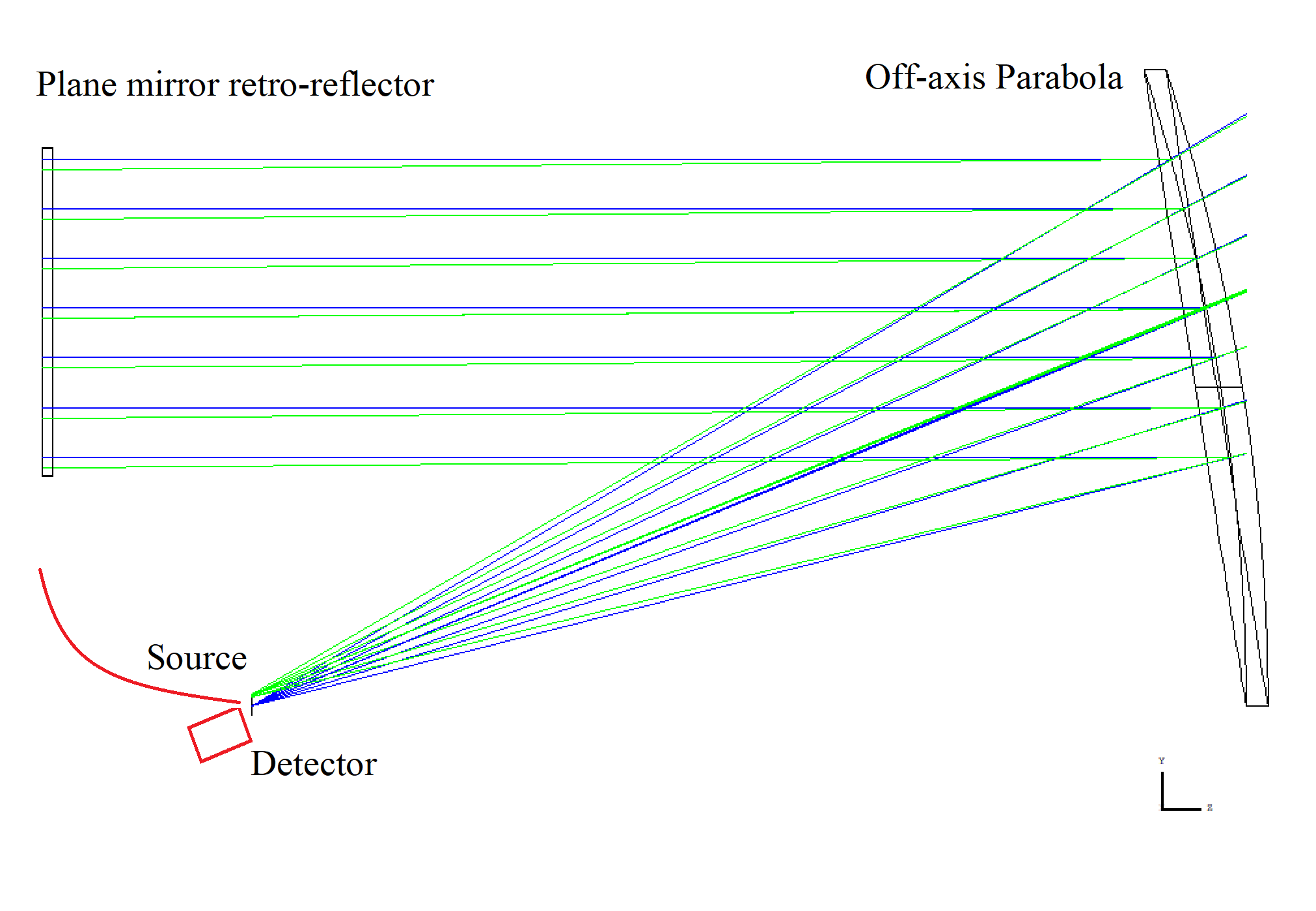}
		\caption{\label{fig:Metro} Schematic of auto-collimation metrology experiment.}
	\end{center}
\end{figure}

\subsection{Metrology Embedded in Telescope}
\label{sec:MET}
Metrology is being used in the Gaia Basic Angle Monitoring (BAM) 
device\cite{Riva2014} 
to keep track of instrument perturbations, small with respect to the scale of 
conventional imaging quality, but relevant to the astrometric error budget. 
The BAM noise performance was specified at the level of a few $\mu as$ over 
the timescale of a few minutes. 
In the case of Gaia, the conceptual requirement is to measure the angle (or 
its variation) between the LOS of two telescopes. 
\\ 
In principle, the LOS is affected by perturbations of any parameter of the optical 
system, although by different amounts which can be estimated by sensitivity analysis. 
A beam combiner removes, at first order, the basic angle issue from the telescope 
and embeds it into a separate subsystem, which might have simple geometry and thus 
hopefully ensure adequate stability thanks to robust mechanical implementation. 

The passive stability approach was sufficient for Hipparcos, designed with a precision 
goal in the $0.1\,milli-arcsec$ ($mas$) range, but it was no longer adequate for Gaia, 
aiming at the $10\,\mu as$ range. 
Given the opto-electrical response variation over the field of view, the mapping 
between position on the focal plane and angular position on the celestial 
sphere is a complex relationship, whose non-linearity must be calibrated 
according to the precision goal. 
This might be accomplished with purely astronomical methods, i.e. by calibration 
on the sky, but the rationale for metrology consists in alleviating the related 
cost, in terms of information, allocated to the stars, thus maximising the scientific 
return. 

Metrology photons should, ideally, follow exactly the optical path of stellar 
photons; in practice, this is only partially possible, and a mismatch will be 
present, on entrance pupil, focal plane, and/or the footprint at intermediate 
optical elements. 
In Gaia, a complex optical setup is required to inject reference laser beams into 
the two telescopes without introducing external errors. 
In a single telescope system, the telescope itself can be used in a sort of 
auto-collimation scheme\cite{Gai12Met,Gai15Met}, as depicted in 
Fig.\,\ref{fig:Metro}: photons are injected close to the focal plane, and the 
return beam is re-imaged in the conjugated position, after a back-and-forth 
travel through the optical system thanks to a retro-reflecting mirror. 
Notably, the back-and-forth travel may compensate or double the disturbances 
on different degrees of freedom; this will be turned to advantage in the 
design of a simple demonstration setup, and modified accordingly for a practical 
implementation on a real telescope, e.g. with multiple beams to disentangle 
the relevant parameters. 

This approach is being tested in OATo lab; the choice of an off-axis parabola 
to represent the telescope, although more critical with respect to alignment, 
allows an easier separation of the source/detector from the retro-reflector. 
The latter region can be exploited for insertion of controlled perturbations 
to the optical paths, simulating the desired disturbances. 

\subsection{Multiple Line Of Sight Telescope}
\label{sec:MLOST}
%
%
The beam combiner concept implemented by Hipparcos was based on two 
half-mirrors cemented together. 
This defines a large angle on a plane, i.e. a one-dimensional angle. 
Simultaneous observation along three, non-planar LOS can similarly be based 
on three, non-redundant angles, providing an operational definition of a 
bidimensional angular measurement on the celestial sphere. 

\begin{figure}[ht]
  \begin{center}
    \includegraphics[width=0.5\textwidth]{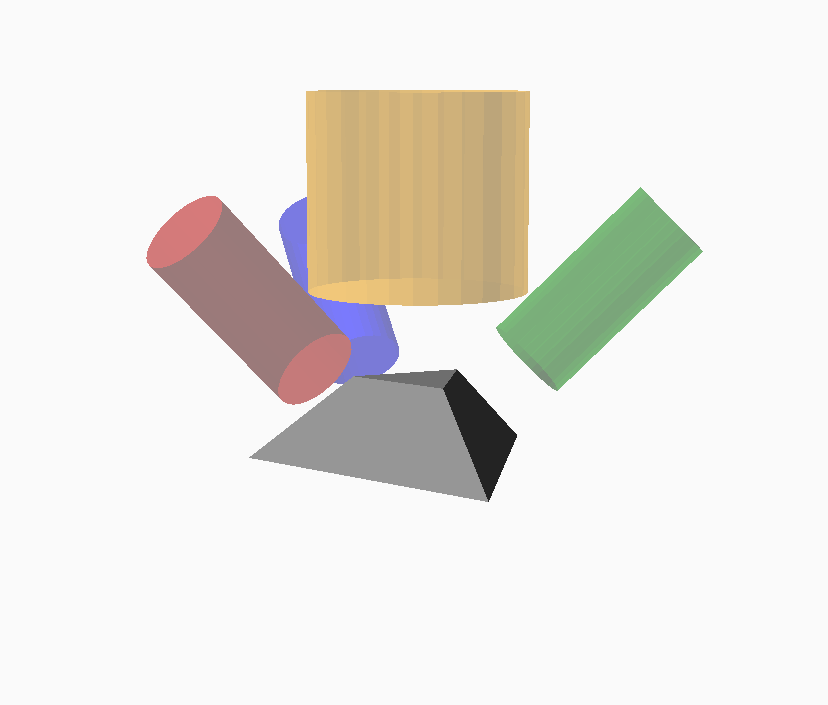}
    \includegraphics[width=0.4\textwidth]{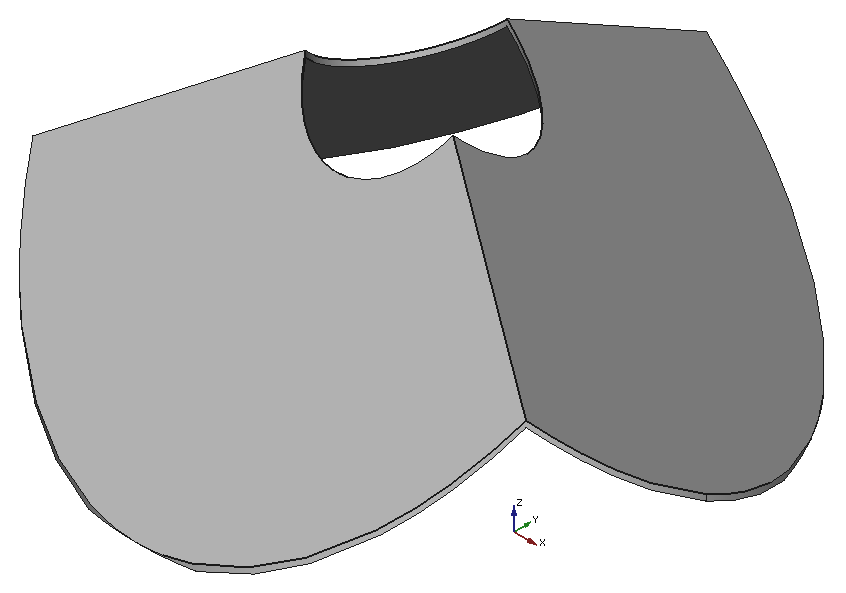}
    \caption{\label{fig:Pyr1} Pyramidal beam combiner principle (left), and a possible 
    lightweight implementation by means of a three mirror assembly.}
  \end{center}
\end{figure}

\begin{figure}[ht]
	\begin{center}
		\includegraphics[width=0.6\textwidth]{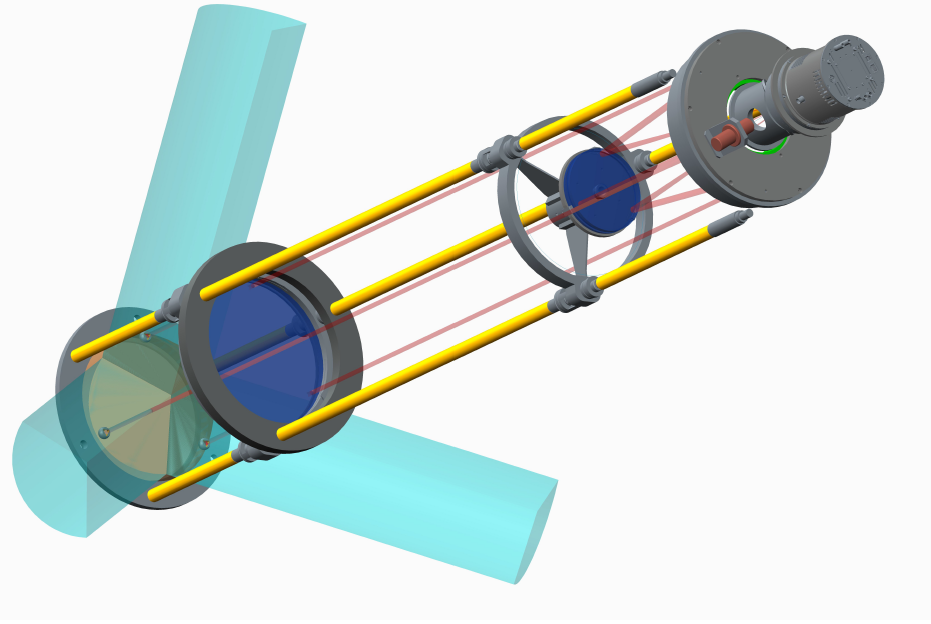}
		\caption{\label{fig:ASTRA_TS} The ASTRA telescope, including the pyramid beam 
			combiner and auto-collimation metrology.}
	\end{center}
\end{figure}

The pyramid approach shown in Fig.\,\ref{fig:Pyr1} (left) provides a simple geometry, 
in which the three lateral faces of the pyramid feed the overall telescope (yellow) 
entrance beam from the incoming photons independently coming from three directions, 
shown respectively in red, blue and green. It may be noted that, reversing the optical 
paths, i.e. shining a light from the focal plane region, collimated by the telescope 
optics, the pyramid acts as a wavefront-splitting beam splitter, sending three beams 
of photons toward the corresponding LOS. 
In figure, the central area has been left free, removing the pyramid peak, e.g. to 
allocate the telescope secondary mirror in a centered system. 

This solution may be materialised by a robust and well defined sub-system, e.g. the 
assembly of three planar mirrors shown in Fig.\,\ref{fig:Pyr1} (right), representing 
a lightweight device, having removed the pyramid bulk. 

The beam combiner represents a convenient location to anchor metrology fiducial points 
(e.g. retro-reflectors) to monitor the relative LOS stability and/or the telescope 
optical response. 
A pictorial representation of the ASTRA telescope, a small size implementation of 
several of the concepts investigated in this study and planned to be implemented 
for forthcoming system-level tests, is shown in Fig.\,\ref{fig:ASTRA_TS}.

\section{CONCLUSIONS}
\label{sec:conclusions}  
The concepts under investigations are considered to be applicable, under appropriate 
rescaling, to the implementation of future high precision astrometry experiments. 
Application examples are presented in other contributions to these proceedings 
(Riva et al., Gai et al.). 
\\ 
Image centering demonstrates the capability of CMOS and CCD detectors to achieve 
a nominal precision corresponding to a small fraction of a pixel ($<1/500$\,pixel 
even in the current initial tests), very close to photon limited expectations. 
\\ 
The design and analysis of a metrology system embedded in a multiple LOS telescope 
is identifying convenient solutions. 
Our roadmap considers as the next step the specifications for a small scale 
experiment aimed at in-flight demonstration of multiple LOS, high precision 
($< 1 \, mas$) astrometry, in preparation for a future implementation on a $1\,m$ 
class telescope devoted to large angle, $\mu as$ level measurements for 
astrophysics and fundamental physics.

\acknowledgments          
 
The INAF activity has been partially funded by a grant from the Italian Ministry of 
Foreign Affairs and International Cooperation, and by the Italian Space Agency 
(ASI) under contracts 2014-025-R.1.2015 and 2018-24-HH.0. 
RAM acknowledges support from CONICYT/FONDECYT Grant Nr. 1190038 and from the 
Chilean Centro de Excelencia en Astrofisica y Tecnologias Afines (CATA) BASAL 
PFB/06.

\bibliographystyle{spiebib}   
\bibliography{mybibl}

\end{document}